\documentclass[12pt]{article}
\usepackage[utf8]{inputenc}
\usepackage[margin=1in]{geometry}

\usepackage{graphicx}
\usepackage{natbib}

\usepackage{amsmath, amsthm, amsfonts, amssymb, latexsym, mathtools}
\usepackage[shortlabels]{enumitem}
\usepackage{bm}
\usepackage{fancyvrb}
\usepackage{longtable}
\usepackage{eurosym}
\usepackage{subcaption}
\usepackage{color}
\usepackage{url}
\usepackage{lscape}
\usepackage{mathrsfs}
\usepackage{enumitem}
\usepackage{amsthm}

\usepackage{tikz}
\usepackage{pgfplots}
\pgfplotsset{compat=1.18}
\usetikzlibrary{quotes,angles}
\usetikzlibrary{shapes,snakes}

\usepackage[english]{babel}
\usetikzlibrary{babel}

\usepackage{booktabs}
\usepackage{caption}
\usepackage{setspace}
\newtheorem{theorem}{Theorem}

\newtheorem{corollary}{Corollary}
\newtheorem{definition}{Definition}

\begin{document}
		
\title{Exploitation, class income and welfare in a Sraffian economy with consumption preferences}
	
\author{Gabriel Montes-Rojas\thanks{Universidad de Buenos Aires-Facultad de Ciencias Económicas, CONICET, 
and Universidad Argentina de la Empresa (UADE), Instituto de Economia (INECO), Ciudad Autónoma de Buenos Aires, Argentina. 
gabriel.montes@economicas.uba.ar} }
	
\maketitle

\begin{abstract}
\noindent This paper develops a Sraffian model of simple reproduction in which production prices and functional income distribution remain classically determined, while class consumption is chosen optimally at those prices and incomes. it then shows that once workers and capitalists choose consumption bundles according to class-specific preferences, rather than consuming fixed baskets, surplus value, exploitation rates, total profits, and capitalist welfare need not rise monotonically with the profit rate.  The framework defines distributive Engel curves, tracing how consumption, profits, labor exploitation, and welfare vary along the wage--profit frontier. Higher profitability may expand capitalists’ nominal claim on surplus, yet alter relative prices in ways that reduce their real consumption and welfare. The result qualifies standard Marxian and Sraffian accounts of net product distribution based on the wage--profit curve.
\end{abstract}
	
	
\pagenumbering{arabic}
		
\newpage

\section{Introduction}

In the simple reproduction Sraffian model, the wage-profit rate $(w,r)$ curve plays a key role to determine how to distribute net product between workers and capitalists. This paper studies class-based preference consumption choices and evaluates how other distributive variables, such as surplus value and exploitation rates, total profits and utility welfare vary along this curve.
It aims to show that these may not be globally (i.e. for all $r$) monotonic, and study necessary and sufficient conditions for this monotonicity.

This analysis contributes to a recent literature that reconsiders the role of demand, household choice, and consumption in classical, Marxian and Sraffian theory. In particular, it builds on \cite{Johansen1963}, \citet{Parchure2019}, \citet{OpocherSteedman24}, \citet{Agata25}, and \citet{Agata24}, who show that preferences and demand can be introduced into the classical framework without reducing prices and distribution to marginal productivity or subjective utility. This integration is important because it clarifies both the scope and the limits of standard neoclassical demand reasoning in long-period positions. As \citet{OpocherSteedman24} argue in their analysis of ``full household equilibrium,'' household income and relative prices cannot be treated as independent variables once they are embedded in a system of production prices. Hence, a change in the distributive variable may generate demand responses that differ sharply from those predicted by partial-equilibrium intuition.

The present paper extends this insight to a class economy. Unlike models based on factor endowments, the economy considered here is divided into workers and capitalists. Workers receive wages, while capitalists receive profits generated in production. Total employment is normalized, and no scarce primary factor other than labor is introduced. Given a profit rate $r$ and wage numeraire $w=1$, production prices are determined in the usual Sraffian manner. Workers  and capitalists  then choose consumption bundles by solving standard utility-maximization problems subject to their respective class incomes. This construction allows us to define \textit{distributive Engel curves}, the paths
which describe how workers' and capitalists' consumption bundles vary as the profit rate changes.\footnote{We depart from the restrictive assumption of a fixed subsistence basket. Following \citet{Salvadori00}, \citet{Signorino01}, and \citet{Stirati91}, the real wage is understood as shaped by social, historical, and institutional conditions rather than by purely physiological requirements. Workers' consumption is therefore modeled as a choice over feasible bundles, not as an exogenously fixed vector. This does not eliminate exploitation or class conflict. On the contrary, it permits a more general analysis of the value of labor-power and of the exploitation rate when the workers' consumption bundle changes with distribution and prices.}

The central result is that the effect of a higher profit rate on capitalist outcomes is not necessarily monotonic in $r$. In the one-consumption-good case and for fixed consumption bundles, the result is unambiguous: as $r$ rises, workers' consumption falls, surplus value rate increases, and capitalists' consumption rises with subsequent increments in total profits, welfare and exploitation. With more than one consumption good and varying consumption bundles, however, the conclusion changes. A variation in $r$ affects class consumption through two channels. First, it redistributes income between workers and capitalists. Second, it changes relative prices. For workers, the income effect is dominant in the sense that welfare declines as the profit rate rises. For capitalists, by contrast, the two effects may move in opposite directions. A higher profit rate may increase their nominal claim on the surplus while simultaneously making their preferred consumption goods more expensive. If this price effect is sufficiently adverse, capitalist real consumption, monetary income (i.e. total profits) and welfare may decline even as the profit rate rises.

This possibility yields a paradoxical implication. The collective pursuit of a higher profit rate need not increase either surplus value, total profits, (which coincides with capitalist expenditure in a simple reproduction economy), or capitalist welfare. In the numerical example developed below, an increase in $r$ changes relative prices in such a way that workers shift consumption toward goods with higher labor values. This reduces the residual labor value available for capitalist consumption, i.e. exploitation reduces. As a result, total profits and capitalist utility may fall over an interval of the profit-rate domain. The mechanism does not deny the classical conflict between labor and capital over the surplus. Rather, it adds a further tension internal to the capitalist class itself: the profit rate that maximizes capitalists' distributive power need not be the profit rate that maximizes their total profits or welfare.

This interpretation is consistent with the classical and Marxian view that consumption is not an autonomous sphere standing outside production. In the \textit{Grundrisse}, Marx argues that production not only supplies the object of consumption but also shapes the manner of consumption and the consuming subject itself: production `produces not only the object but also the manner of consumption'' and thereby `creates the consumer'' \citep[pp.~92--93]{marx1973grundrisse}. He later summarizes the point by stating that `production predominates \ldots\ over the other moments as well'' and that `a definite production thus determines a definite consumption, distribution and exchange'' \citep[p.~99]{marx1973grundrisse}. Similarly, in \textit{Capital}, Volume II, consumption first appears within the circuit of capital as productive consumption, that is, as a moment of reproduction rather than as an independent determinant of the system \citep[p.~197]{marx1993capital2}. The approach adopted here follows this ordering: distribution and prices are determined prior to final consumption choices, but those choices remain essential for understanding class-specific welfare.

The analysis also relates to Kalecki's proposition that ``capitalists, as a whole, determine their own profits by the extent of their investment and personal consumption'' \citep[p.~13]{Kalecki71}. In the present framework, capitalist consumption equals total profits in equilibrium. Yet the amount of capitalist consumption compatible with reproduction depends on prices, techniques, and the composition of workers' consumption. Thus, capitalist profits are not merely a mechanical increasing function of the profit rate. Once the composition of consumption is allowed to vary, the monetary and welfare consequences of a higher profit rate may diverge.

The paper has two main implications. First, it provides a formal tool for studying how distributive changes affect class-specific consumption: the distributive Engel curve. Second, it shows that the maximum technically feasible profit rate $R$, determined by the input--output structure, need not be the relevant upper bound from the standpoint of capitalist welfare. If capitalist welfare reaches a maximum at some interior value $r^*<R$, then further increases in the profit rate may be technically feasible but income and welfare-reducing for capitalists themselves. The traditional conflict between workers and capitalists over surplus distribution is therefore complemented by a possible contradiction within the capitalist class: the pursuit of a higher profit rate may undermine capitalists' own consumption possibilities.

The paper is organized as follows. Section \ref{sec:eqgral} presents the general model of production prices with class-specific consumption preferences and defines distributive Engel curves, profits, and welfare. Section \ref{sec:dosbienes} studies the multi-good case and provides a numerical example in which capitalist profits and welfare decline as the profit rate rises. It also derives necessary and sufficient conditions for the monotonicity of total profits and capitalist welfare. Section \ref{conclusion} concludes with implications for the political economy of distribution and class conflict. Mathematical proofs are collected in the Appendix.

\section{Consumption preferences in the classical model}\label{sec:eqgral}

\subsection{Prices of production with consumption preferences}

Consider a simple reproduction economy with only circulating capital and $n$ sectors, each producing one commodity. Let $ A=[a_{ij}]$ be the $n\times n$ input-output matrix, where $a_{ij}$ denotes the quantity of commodity $i$ needed to produce one unit of commodity $j$. Assume $A$ satisfies the Hawkins-Simon conditions (see \citet[][Ch. 4]{Takayama85}). In particular we assume that it is nonnegative, productive, and irreducible. The model follows the standard prices of production scheme presented in \cite{KurzSalvadori95} and \cite{bidard04}.

Let $x=[x_1,\ldots,x_n]'$ be the $n\times 1$ vector of gross outputs and $ c=[c_1,\ldots,c_n]'$ the vector of final consumption. Then
\[
 A x +  c =  x,
\]
so that $ x=( I- A)^{-1} c$, provided $( I- A)$ is nonsingular.

The model must satisfy two standard conditions. First, viability requires that consumption does not exceed net output. In the absence of investment,
\begin{equation}\label{eq:viabilidad}
 c = ( I- A) x.
\end{equation}
Second, since labor is the only primary factor, we normalize total employment:
\begin{equation}\label{eq:plenoempleo}
 \ell  x = 1,
\end{equation}
where $ \ell=[\ell_1\; \ell_2 \;...\; \ell_n]\geq0$, a $1\times n$ vector, represents labor requirements per unit output.

Prices follow the standard \cite{Sraffa60} model of production prices. Using $w=1$ as numeraire, the main distributive variable is $r$, the uniform profit rate, such that
\begin{equation}
     p  A(1+r)+ \ell= p,
\end{equation}
where $ p=[p_1,\dots,p_n] \geq  0_n$ is the vector of production prices. Prices $ p(r)$ are defined for $r \in [0,R)$, where $R = 1/\rho( A) - 1$ and $\rho( A)$ is the Perron root of $ A$. At $r=R$, we have $ p(R) A(1+R)= p(R)$, while at $r=0$, $ p(0)= v$, where $v$ is the vector of labor values defined below, see eq. \ref{eq:valores}. Thus,
\begin{equation}\label{eq:p}
     p(r) =  \ell \left(I-(1+r) A\right)^{-1}, \quad r \in [0,R).
\end{equation}

Total consumption is the sum of workers' and capitalists' consumption:
\begin{equation}
 c =  c^\ell +  c^k.
\end{equation}

We assume a representative worker-consumer with preferences $\succeq^\ell$ represented by a utility function $U^\ell:\mathbb{R}_+^n \to \mathbb{R}$. The worker's problem is
\begin{equation}\label{eq:ul}
    \max_{ c^\ell} U^\ell( c^\ell) \quad \text{subject to }  p c^\ell \leq  \ell  x \quad \text{for } r \in [0,R).
\end{equation}
The solution yields Marshallian demands $ h^\ell( p,1)$ with workers' income being $ \ell x = 1$, or indexed by the profit rate, $c^\ell(r):=h^\ell( p(r),1)$. For $r=R$, workers have no income, so $ c^\ell(R)= 0$.

Similarly, the representative capitalist-consumer with preferences $\succeq^k$ solves
\begin{equation}\label{eq:uk}
    \max_{ c^k} U^k( c^k) \quad \text{subject to }  pc^k\leq m  \quad \text{for } r \in (0,R],
\end{equation}
where $m$ is capitalists' income and with solution being a Marshallian demand $c^k(r):=h^k(p(r),m(r))$, where $m(r)$ is capitalists income that may vary with $r$. For $r=0$, capitalists have no income, so $ c^k(0)= 0$.

We assume that preferences are `well behaved', such that the results are not driven by specific assumptions. In particular,
for \(C=\mathbb{R}^n_{+}\), the preference relations \(\succeq\) for $\ell$ and $k$ on \(C\) are
rational, continuous, locally non-satiated, convex and monotone.  That is, \(\succeq\)
is complete and transitive; its upper and lower contour sets are closed;
for every \(c\in C\) and every \(\varepsilon>0\), there exists
\(y\in C\) with \(\|y-c\|<\varepsilon\) such that \(y\succ c\); 
for all \(c,y\in C\) and all \(\lambda\in[0,1]\), if \(c\succeq y\), then
\(\lambda c+(1-\lambda)y\succeq y\); and for
all \(c,y\in C\), if \(y\geq c\), then \(y\succeq c\). We also impose uniqueness of the consumer's
optimal choice such that \(\succeq\) is assumed to be strictly convex.
These are standard assumptions in classical demand theory
\cite[Ch.~3, especially Sec.~3.B]{MasColell95}.

We consider an adaptation of the full household equilibrium as in \cite{OpocherSteedman24} and a general equilibrium model as in \cite{Parchure2019} that combines the Sraffian prices of production scheme with consumption preferences. This is a special case of \cite{Johansen1963} model where we fix $r$.

\begin{definition}\label{def:longrun}
\textit{Long-run position with consumption preferences.} With $w=1$ as numeraire and a fixed value of $r\in[0,R]$, the long-run position with consumption preferences is given by:
\begin{itemize}
    \item[(A)] a gross product vector $ x^*$ and aggregate consumption $ c^*$ satisfying \eqref{eq:viabilidad} and \eqref{eq:plenoempleo};
    \item[(B)] a price vector $ p^*$ and profit rate $r^* \in [0,R]$ satisfying \eqref{eq:p};
    \item[(C)] $ c^* =  c^{\ell*} +  c^{k*}$, where $ c^{\ell*}$ solves \eqref{eq:ul} and $ c^{k*}$ solves \eqref{eq:uk}.
\end{itemize}    
\end{definition}

Condition (A) captures supply–demand equilibrium, (B) determines prices and the profit rate given exogenous distribution, and (C) adds optimal consumption behavior. 

Here we consider that capitalist-consumer takes prices and profits as given, and their consumption decisions are independent of the production process. In this case, we consider that the production sphere is prior to the circulation process where consumption takes place.

Then
\begin{equation}
    m(r)=rp(r)Ax(r)
\end{equation}
 is capitalists' income taking production prices and total product as given. Here consumption decisions are based on a given Marshallian demand at the prices of production $p(r)$.
Note that this model implies that in equilibrium

\begin{equation}
    p(r) c^{k*}(r)=p(r) h^k(p(r),m(r))=m(r)=rp(r)Ax^*(r),
\end{equation}
such that capitalists' consumption equals total profits. Then,

\begin{equation}
    p(r) c^{k*}(r)=rpA(I-A)^{-1}(c^{\ell*}(r)+c^{k*}(r))
\end{equation}




\subsection{Labor value, surplus value and exploitation}

Labor values $v=[v_1\; v_2 \;...\; v_n]\geq0$ are a $1\times n$ vector  obtained from
\[
v  A + \ell = v,
\]
yielding
\begin{equation}\label{eq:valores}
 v = \ell(I- A)^{-1}.
\end{equation}
Multiplying both sides of $x=(I-A)^{-1}c$ by $\ell$ gives $\ell x = vc = 1$, so
\begin{equation}\label{eq:rest}
vc = 1,
\end{equation}
which must hold for any income distribution. This determines that

\begin{equation}
    vc^\ell(0) = vc^k(R)  = 1
\end{equation}
where either class consumes all net product.

Let $\sigma$ be the surplus value and let $c^\ell$ be the worker's consumption wage bundle, a $n\times 1$ vector. Given the value vector $v$, the surplus value rate $\sigma(r)$ (indexed by $r$) can be determined from the equation:

\begin{equation}\label{eq:surplus}
(1+\sigma(r))vc^\ell(r) =1 ,
\end{equation}
\begin{equation}\label{eq:surplus2}
\sigma(r)=\frac{1}{vc^\ell(r)}-1=\frac{1-vc^\ell(r)}{vc^\ell(r)} ,
\end{equation}
where $vc^\ell(r)$ is the value of labor force. In the standard Marxist model this consumption bundle is given. Here we emphasize that this is a ``choice'' made by workers and not a given set of goods.

Using $vc(r)=1$ by \eqref{eq:rest}, together with \eqref{eq:surplus2}, determines that 

$$\sigma(r)=\frac{vc^k(r)}{vc^\ell(r)},$$
that is, the surplus rate is the ratio of values of capitalists' consumption to workers' consumption. Here we defined $v c^a(r)$, $a=\ell,k$ as the value of consumption bundles by capitalists and workers, respectively.

For convenience, consider the the exploitation rate as

$$e(r):=1-v c^\ell(r).$$
Note that $e$ is a monotonic transformation of $\sigma$ but it has the advantage that is in the $[0,1]$ closed interval, whereas $\sigma\in[0,\infty)$. Here it is clear that

\[e(0)=0\]

\[e(R)=1\]

However, the behavior of $e(r),\ r\in(0,R)$ may be non monotonically increasing in $r$. The reason is that changes in prices may change the labor content of the consumed commodities. As such, we may have the case that the exploitation rate or the surplus value rate may decrease when $r$ increases.

For further reference, let \[
H(r,m):=v h^k(p(r),m).
\]
be the labor value of the capitalist bundle purchased with income \(m\). Here $r$ only affects prices, and then $H_r$ is the derivative of $H$ with respect to the first component only, a value price substitution effect, while $H_m$ is the derivative with respect to the second component, a value income effect. 

The equilibrium capitalist expenditure \(m(r)\) is determined by the condition
that capitalist consumption absorbs the residual labor value:
\[
H(r,m(r))=e(r).
\]

\subsection{Distributive Engel curves, profit and welfare}

We define
\[
( c^\ell(r), c^k(r)),\quad r\in[0,R]
\]
as the \textit{distributive Engel curves}.

Profit functions are defined as
\[
\pi(r)=rp(r)Ax(r)=p(r)c^k(r),\quad r\in[0,R].
\]

Welfare functions can then be obtained by substituting $ c^\ell(r)$ and $ c^k(r)$ into the corresponding utility functions.

Finally, let
\[
U^\ell(r):=U^\ell( c^\ell(r)),\quad U^k(r):=U^k( c^k(r)),\quad r\in[0,R]
\]
as the \textit{distributive welfare functions}.

\section{Main results}\label{sec:dosbienes}

Consider the case of more than one consumption commodity, so that consumption decisions  may vary when prices and income distribution also vary. Workers and capitalists choose consumption bundles optimally and the solution satisfies Definition \ref{def:longrun}.

First, we develop a full numerical exercise in Subsection \ref{subsec:numerical} to show that a non monotonic case may occur for canonical cases. For this we consider a $n=2$ commodities case with workers CES preferences and capitalists Cobb-Douglas preferences.

Second, we provide necessary and sufficient conditions for $\pi(r)$ (Subsection \ref{subsec:profits}) and capitalist welfare $U^k(r)$ (\ref{subsec:welfare}) to be non-decreasing on $(0,R)$. The conditions are related to labor value conditions based on the exploitation rate $e(r)$ and labor value of capitalists consumption at fixed income, $H(r,m)$.

Third we consider the special case of fixed consumption bundles in Subsection \ref{subsec:fixed}.

\subsection{Numerical exercise}\label{subsec:numerical}
We provide here a numerical exercise where profits, capitalists' welfare function and exploitation rate can be decreasing on the $(0,R)$ interval.

Consider a two-commodity circulating-capital economy with input-output matrix and unit labor requirements
\[
A=
\begin{pmatrix}
1/5 & 7/25\\
1/20 & 43/100
\end{pmatrix},
\qquad
\ell=\left(\frac32,\frac15\right).
\]
For this case $R=\frac{13}{12}$.

Workers have CES preferences,
\[
U^\ell(c_1,c_2)
=
\left[
\delta c_1^{(\sigma-1)/\sigma}
+
(1-\delta)c_2^{(\sigma-1)/\sigma}
\right]^{\sigma/(\sigma-1)},
\]
with parameters $\delta=0.40$ and $\sigma=20$, and income normalized to one. 

Capitalists have Cobb-Douglas preferences

\[
U^k(c_1,c_2)
=c_1^{\beta}c_1^{1-\beta},
\]
with parameter $\beta=0.8$.

The point of the numerical example is to show that the profits, capitalists welfare and exploitation
can be locally decreasing in \(r\), even though \(r\) itself is the profit rate. In the exercise, this is achieved through a change in workers' demand that makes workers' consumption more labor-intensive over a range of profit rates. Since workers' income is normalized to 1, workers react to a substitution effect, i.e. price change when varying $r$.\footnote{Similar results can be obtained in the simplest case of perfect substitutes preferences for the two goods. When the profit rate changes, relative prices change too. Workers and capitalists may change their consumption bundle in a way that reduce exploitation rate, profits and welfare.  Here we opt to develop an example with continuously differentiable demand functions.}

\begin{figure}[htbp]
    \centering
    \begin{subfigure}{0.48\linewidth}
        \centering
        \includegraphics[width=\linewidth]{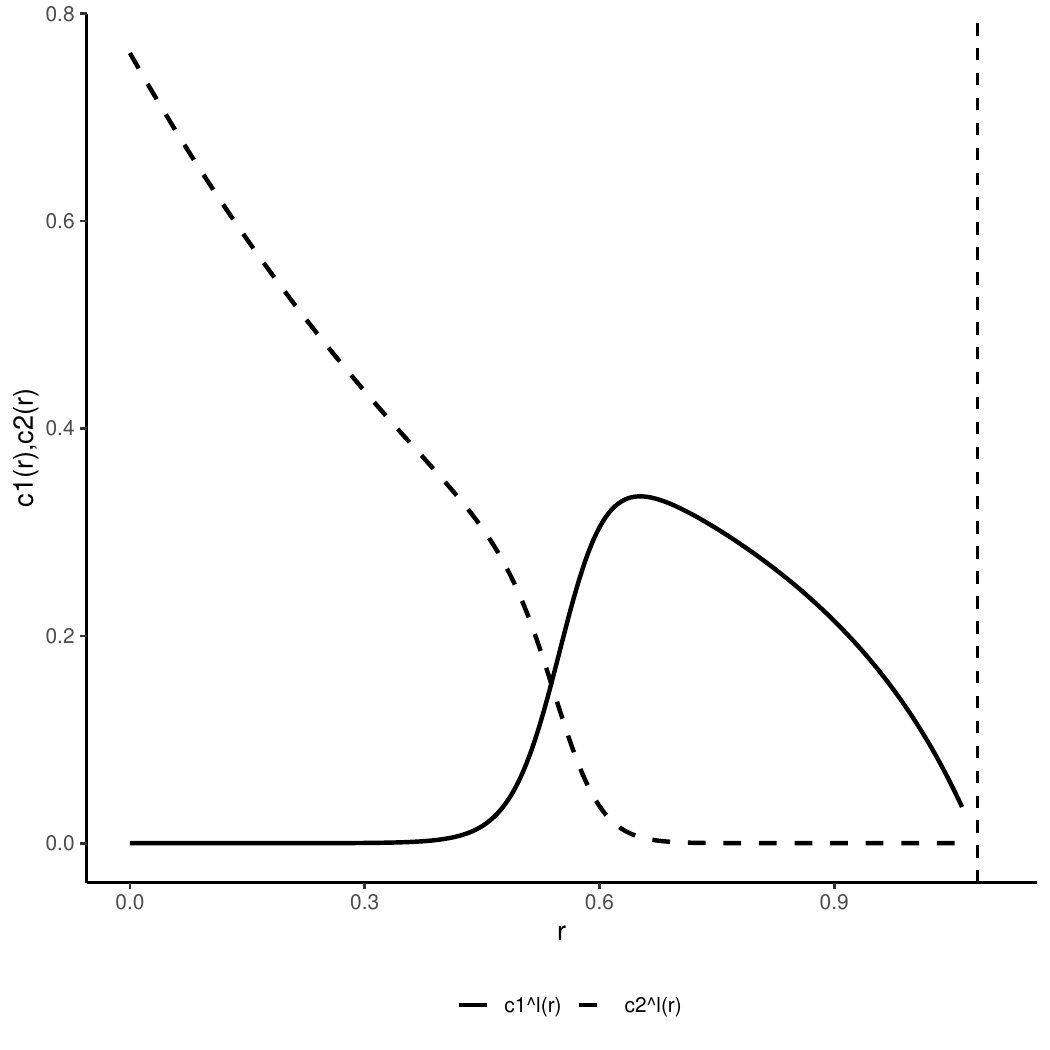}
        \caption{Workers}
        \label{fig:worker_c1c2}
    \end{subfigure}
    \hfill
    \begin{subfigure}{0.48\linewidth}
        \centering
        \includegraphics[width=\linewidth]{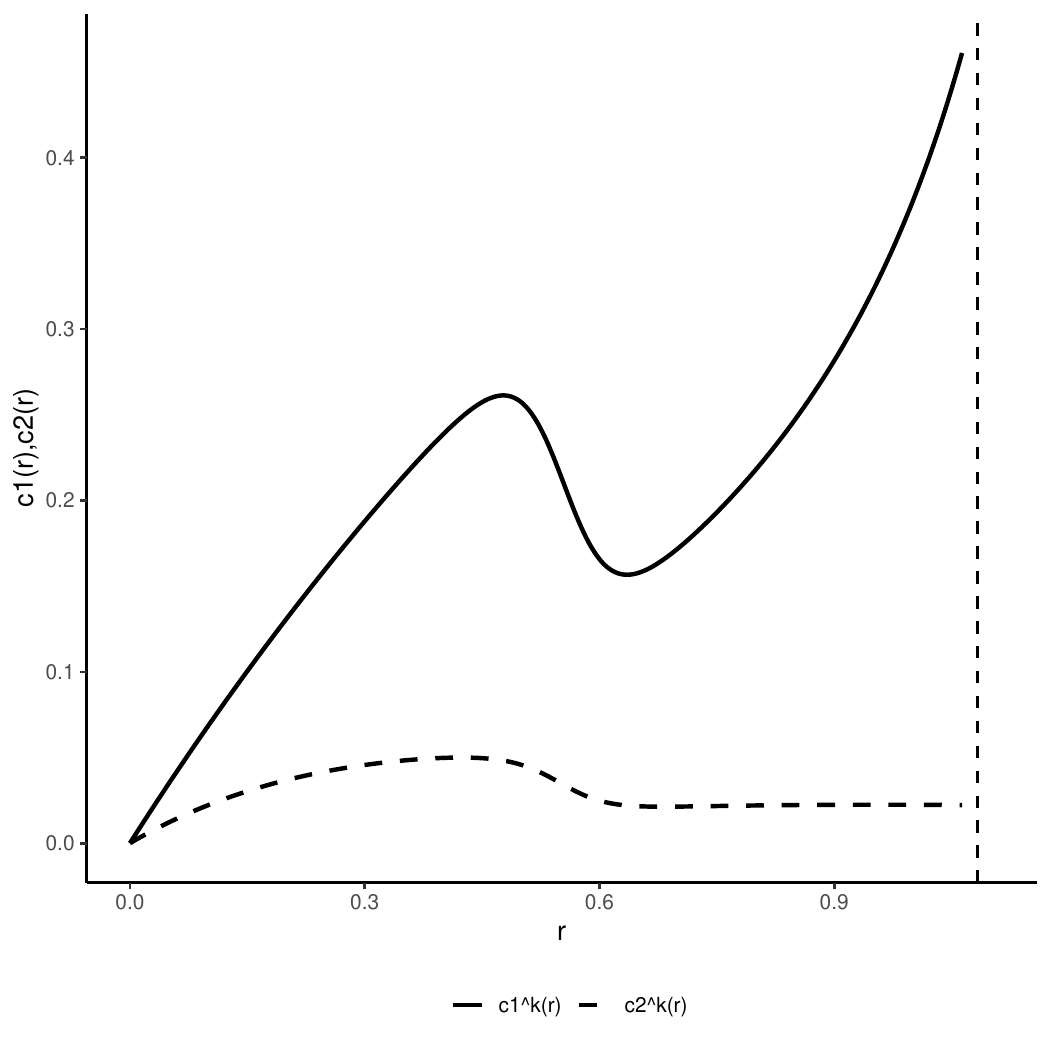}
        \caption{Capitalists}
        \label{fig:cap_c1c2}
    \end{subfigure}

    \caption{Consumption Engel curves $(c_1(r),c_2(r))$}
    \label{fig:c1c2}
\end{figure}

Figure \ref{fig:c1c2} plots the consumption patterns for $r\in(0,R)$. The values of the CES parameters make workers react to changes in relative prices, so that when they start consuming more of $c_1$, capitalists must reduce their own consumption of $c_1$.

\begin{figure}[htbp]
    \centering
    \begin{subfigure}{0.48\linewidth}
        \centering
        \includegraphics[width=\linewidth]{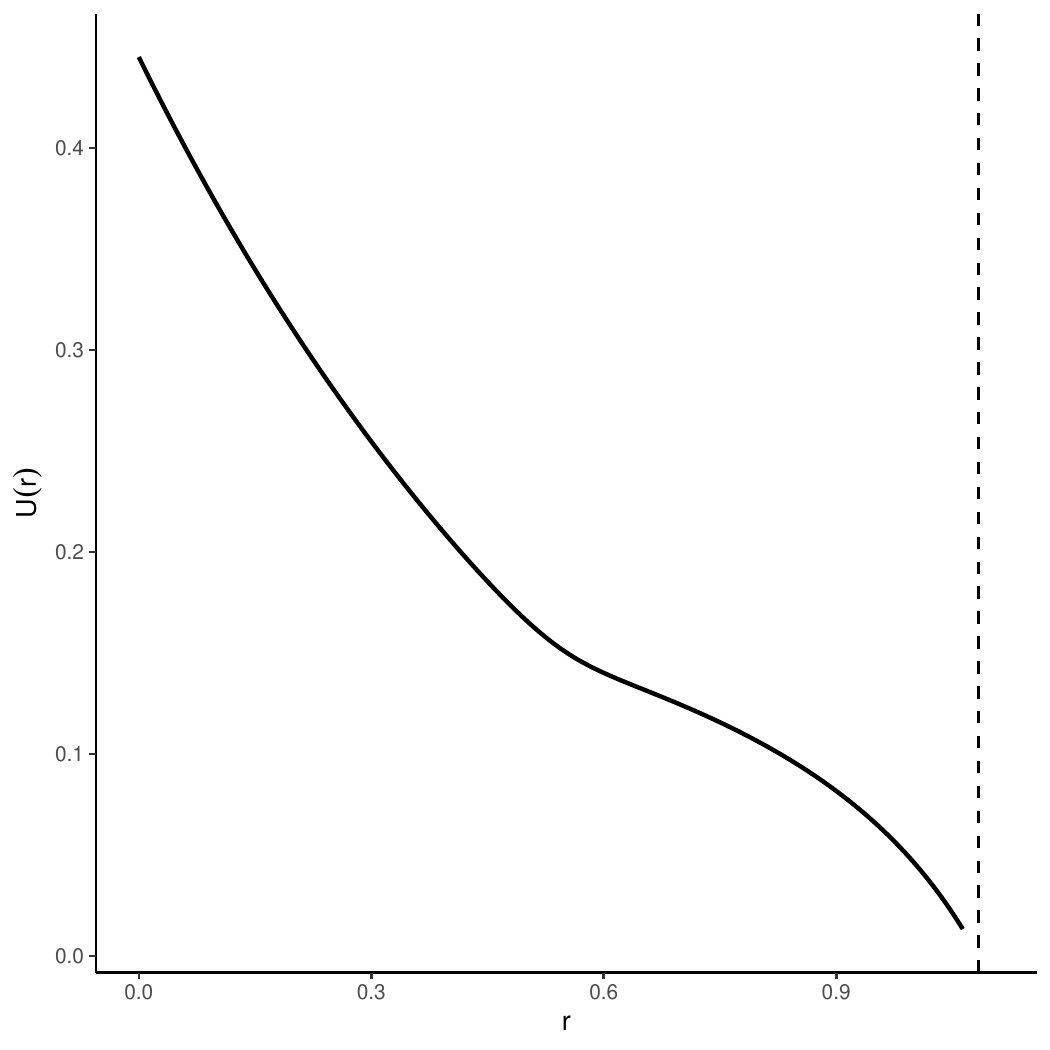}
        \caption{Workers}
        \label{fig:worker_u}
    \end{subfigure}
    \hfill
    \begin{subfigure}{0.48\linewidth}
        \centering
        \includegraphics[width=\linewidth]{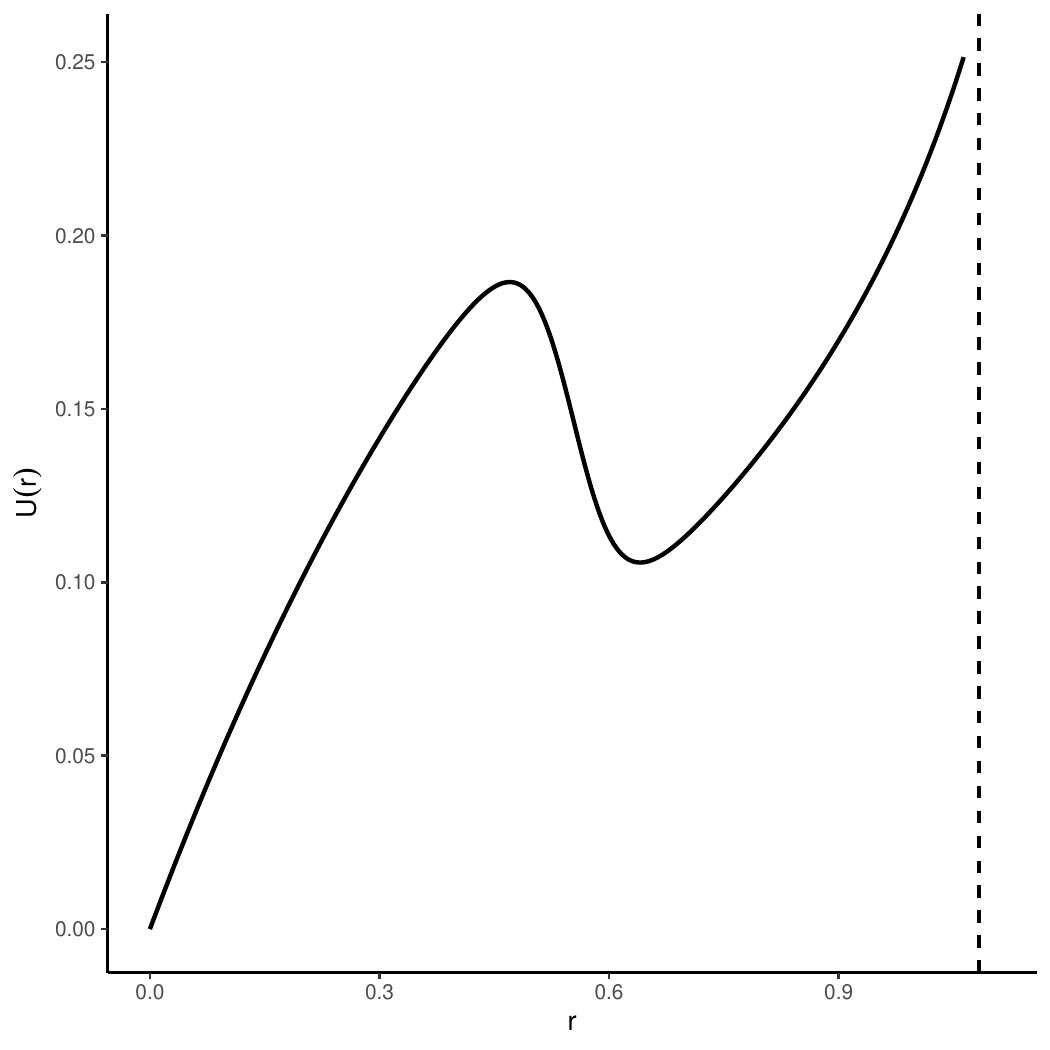}
        \caption{Capitalists}
        \label{fig:cap_u}
    \end{subfigure}

    \caption{Welfare functions}
    \label{fig:u}
\end{figure}

In turn, this significant shift in consumption produces a reduction of capitalists welfare, as seen in Figure \ref{fig:u}.

\begin{figure}[htbp]
    \centering
    \begin{subfigure}{0.48\linewidth}
        \centering
        \includegraphics[width=\linewidth]{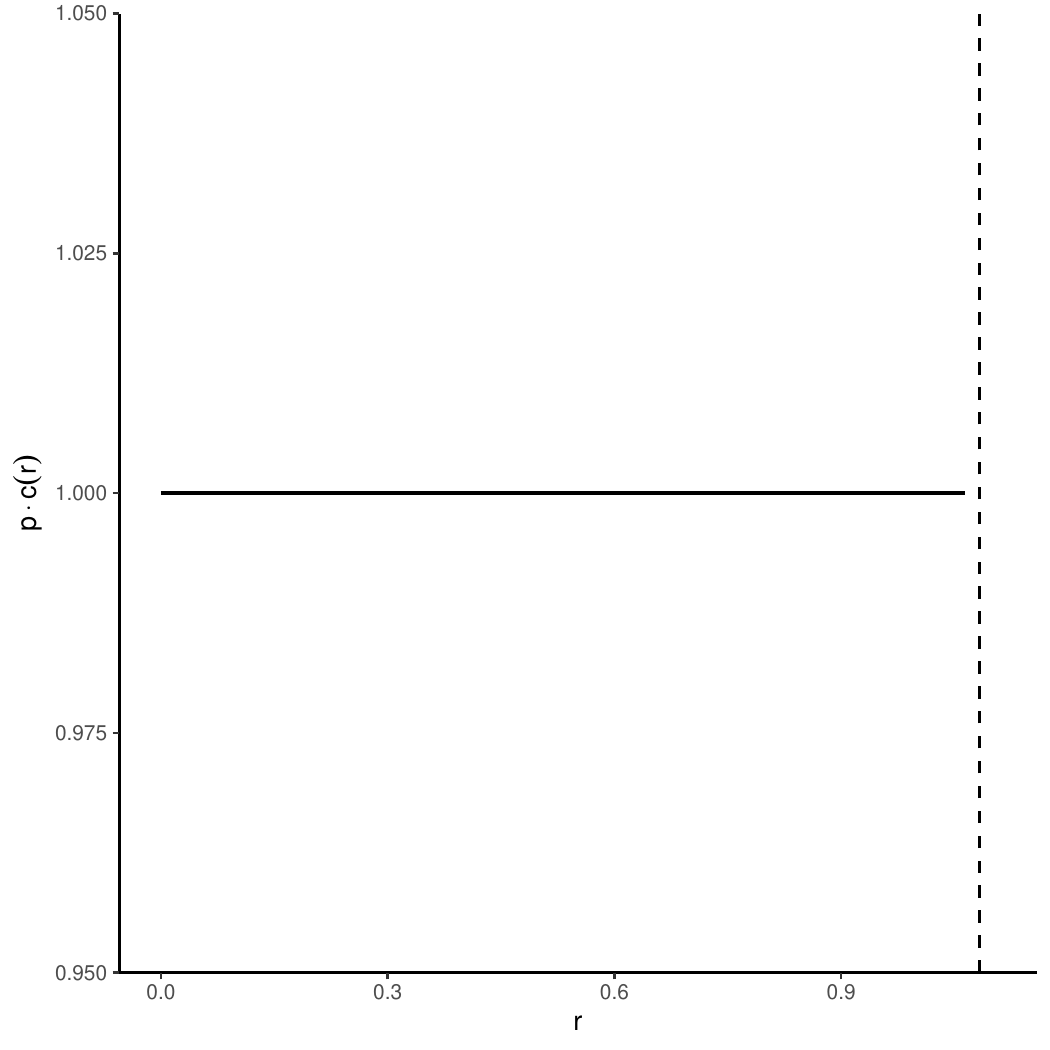}
        \caption{Workers}
        \label{fig:worker_pc}
    \end{subfigure}
    \hfill
    \begin{subfigure}{0.48\linewidth}
        \centering
        \includegraphics[width=\linewidth]{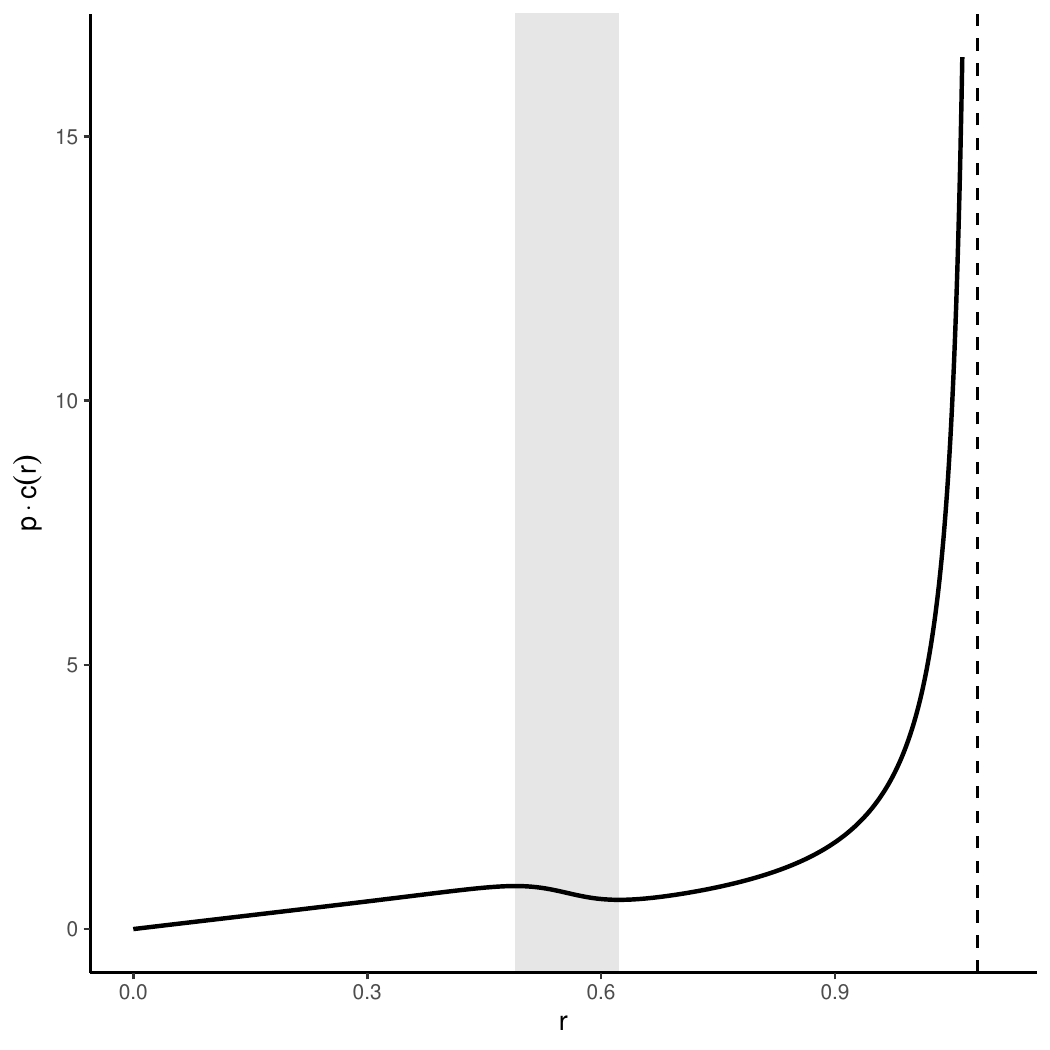}
        \caption{Capitalists $=$ total profits}
        \label{fig:cap_pc}
    \end{subfigure}

    \caption{Income functions}
    \label{fig:pc}
\end{figure}

Moreover, this change in consumption patterns also alters total profits. Figure \ref{fig:pc} shows that for an interval on $(0,R)$ total profits (measured by $p(r)c^k(r)$) decrease as $r$ increases.

\begin{figure}[htbp]
    \centering
    \begin{subfigure}{0.48\linewidth}
        \centering
        \includegraphics[width=\linewidth]{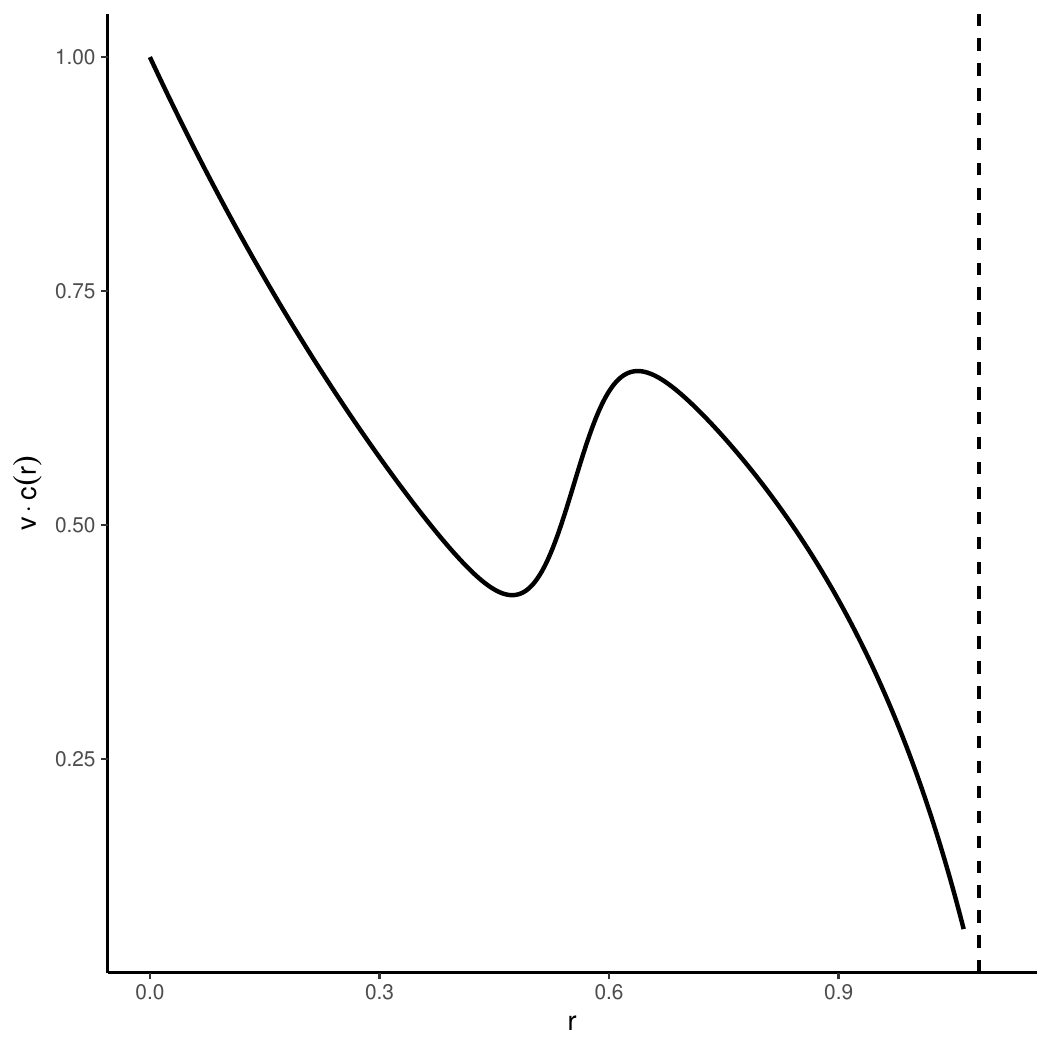}
        \caption{Workers $=$ value of labor force}
        \label{fig:worker_vc}
    \end{subfigure}
    \hfill
    \begin{subfigure}{0.48\linewidth}
        \centering
        \includegraphics[width=\linewidth]{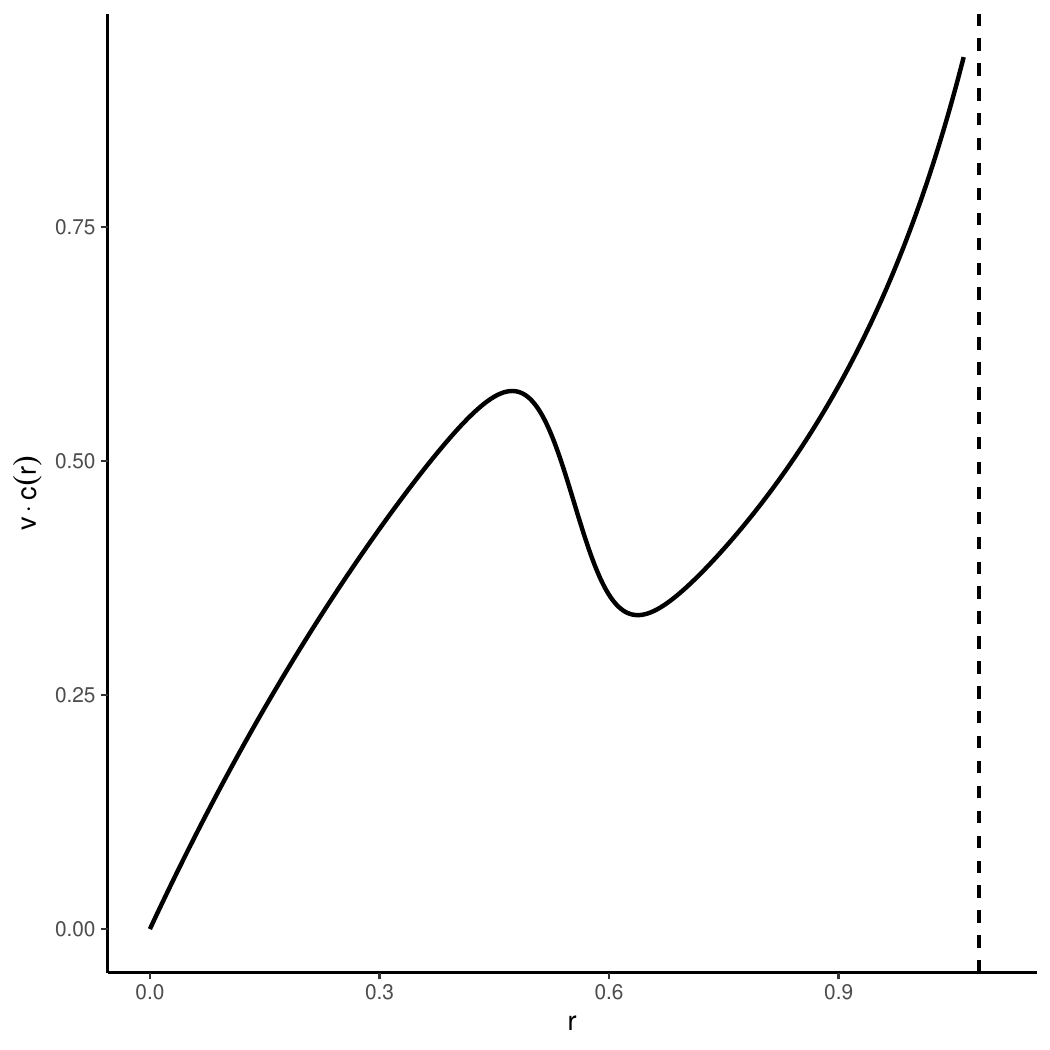}
        \caption{Capitalists $=$ exploitation rate}
        \label{fig:cap_vc}
    \end{subfigure}

    \caption{Value of consumption}
    \label{fig:vc}
\end{figure}

Finally, Figure \ref{fig:vc} shows that the exploitation rate is not monotonic in $r$. Changes in consumption patterns change the labor value composition of workers' consumption, thus changing surplus and exploitation rates.

\subsection{A necessary and sufficient condition for weakly monotonic total profits}\label{subsec:profits}

Here we explore the conditions under which total profits (which equals capitalists consumption) cannot decrease when $r$ increases. The following theorem establishes a clear relation with labor value theory, such that profits are weakly increasing in $r$ whenever change in the exploitation rate change \(e'(r)\) is larger than the labor value change of the price effect \(H_r(r,m(r))\), a substitution type effect.

\begin{theorem}\label{thm:necsuf_profit}[Necessary and sufficient local condition]
Suppose that \(p(r)\), \(h^\ell(p(r),1)\), and \(h^k(p(r),m)\) are continuously
differentiable on the relevant domain on $(0,R)$. Assume that workers and capitalists preferences are globally non-saciated so that they exhaust their budget, in particular,\(p(r)c^\ell(r)=1\) and
\(p(r)c^k(r)=p(r)h^k(p(r),m(r))=m(r)\). Suppose also that for every \(r\)
there exists a unique \(m(r)>0\) satisfying
\(H(r,m(r))=e(r)\).

If $H_m(r,m(r))\neq 0$,
then capitalist expenditure is locally weakly increasing at \(r\) if and only if
$$\frac{e'(r)-H_r(r,m(r))}{H_m(r,m(r))}\geq 0.$$
That is, $$\frac{d}{dr}\bigl[p(r)c^k(r)\bigr]\geq 0
\quad\Longleftrightarrow\quad
\frac{e'(r)-H_r(r,m(r))}{H_m(r,m(r))}\geq 0.$$
In particular, if $H_m(r,m(r))>0$,
then capitalist expenditure is locally weakly increasing at \(r\) if and only if $$e'(r)\geq H_r(r,m(r)).$$
\end{theorem}

Consider here the relation with the labor-content equilibrium equation and the numerical example.

For Cobb--Douglas capitalists, expenditure shares are constant:
\[
s_1^k=\beta,\qquad s_2^k=1-\beta.
\]
Define the labor commanded by one unit of money spent on commodity \(i\) by
\[
S_i(r)=\frac{v_i}{p_i(r)}.
\]
Then the labor commanded per unit of capitalist expenditure is
\[
\bar S^k(r)
=
\beta S_1(r)+(1-\beta)S_2(r).
\]
Therefore,
\[
H(r,m)=m\bar S^k(r).
\]
The labor-content equilibrium condition becomes
\[
m(r)\bar S^k(r)=e(r),
\]
so
\[
m(r)=\frac{e(r)}{\bar S^k(r)}.
\]
This is the same object as \(m(r)=p(r)c^k(r)\), just expressed from the labor-content
side rather than the production-income side.

For Cobb--Douglas capitalists,
\[
H(r,m)=m\bar S^k(r).
\]
Thus
\[
H_m(r,m)=\bar S^k(r)>0,
\]
and
\[
H_r(r,m)=m(\bar S^k)'(r).
\]
The derivative of equilibrium expenditure is therefore
\[
m'(r)
=
\frac{e'(r)-m(r)(\bar S^k)'(r)}{\bar S^k(r)}.
\]
Equivalently, since \(m(r)=e(r)/\bar S^k(r)\),
\[
m'(r)
=
\frac{e'(r)\bar S^k(r)-e(r)(\bar S^k)'(r)}
{\left(\bar S^k(r)\right)^2}.
\]
Hence
\[
m'(r)<0
\quad\Longleftrightarrow\quad
e'(r)\bar S^k(r)<e(r)(\bar S^k)'(r).
\]

This gives the precise reason why profits can be locally decreasing. A rise in \(r\) may
induce workers to shift their consumption toward goods with higher labor values. Then \(vc^\ell(r)\) may rise, causing \(e(r)=1-vc^\ell(r)\)
to fall. If this fall in residual labor value is sufficiently strong, it can dominate any
change in the capitalist labor-commanded expenditure term \(\bar S^k(r)\). In that case \(m'(r)<0\).

Specifically, the residual labor
value \(e(r)\) falls too quickly relative to the fixed-income labor content of capitalist
demand. As a result, the equilibrium capitalist expenditure required to absorb the residual
labor value falls even though the profit rate \(r\) rises.

\subsection{A necessary and sufficient condition for weakly monotonic capitalists welfare}\label{subsec:welfare}

Capitalist welfare along the equilibrium path is
\[
U^k(r):=U^k(c^k(r)).
\]
Equivalently,
\[
U^k(r)=V^k(p(r),m(r)).
\]
Define
\[
W(r,m):=V^k(p(r),m).
\]
Then
\[
U^k(r)=W(r,m(r)).
\]

\begin{theorem}\label{thm:necsuf_welfare}[Necessary and sufficient condition for capitalist welfare] 
Suppose that \(p(r)\), \(h^\ell(p(r),1)\), \(h^k(p(r),m)\), and
\(V^k(p(r),m)\) are continuously differentiable on the relevant domain. Suppose
also that, for every \(r\), there exists a unique \(m(r)>0\) satisfying $H(r,m(r))=e(r)$.

If $H_m(r,m(r))\neq 0$,
then capitalist welfare is locally weakly increasing at \(r\) if and only if $$W_r(r,m(r))
+
W_m(r,m(r))
\frac{e'(r)-H_r(r,m(r))}{H_m(r,m(r))}
\geq 0.$$ That is, $$\frac{dU^k(r)}{dr}\geq 0
\quad\Longleftrightarrow\quad
W_r(r,m(r))
+
W_m(r,m(r))
\frac{e'(r)-H_r(r,m(r))}{H_m(r,m(r))}
\geq 0.$$
Equivalently,
$$
\frac{dU^k(r)}{dr}\geq 0
\quad\Longleftrightarrow\quad
\frac{
W_r(r,m(r))H_m(r,m(r))
+
W_m(r,m(r))\left[e'(r)-H_r(r,m(r))\right]
}{
H_m(r,m(r))
}
\geq 0$$. If, in addition,
$H_m(r,m(r))>0$,
then capitalist welfare is locally weakly increasing at \(r\) if and only if
$$W_r(r,m(r))H_m(r,m(r))
+
W_m(r,m(r))\left[e'(r)-H_r(r,m(r))\right]
\geq 0.$$
\end{theorem}

The condition for capitalist expenditure to be non-decreasing is simply
\[
m'(r)\geq 0.
\]
The condition for capitalist welfare to be non-decreasing is different because prices also
change with \(r\). Welfare changes according to
\[
\frac{dU^k(r)}{dr}
=
\underbrace{W_r(r,m(r))}_{\text{price effect at fixed income}}
+
\underbrace{W_m(r,m(r))m'(r)}_{\text{income effect}}.
\]
Thus capitalist welfare can rise even if capitalist expenditure falls, provided the price
effect is favorable enough. Conversely, capitalist expenditure can rise while welfare falls
if the price effect is sufficiently unfavorable.

\begin{corollary}\label{cor:necsuf_roy}
[Roy identity formulation]
Suppose, in addition, that the indirect utility function \(V^k(p,m)\) is differentiable,
that \(V_m^k(p,m)>0\), and that Roy's identity applies:
\[
\nabla_p V^k(p,m)
=
- V_m^k(p,m) h^k(p,m).
\]
Then capitalist welfare is locally weakly increasing at \(r\) if and only if $\frac{e'(r)-H_r(r,m(r))}{H_m(r,m(r))}
\geq
h^k(p(r),m(r))\cdot p'(r)$. Equivalently,
$\frac{dU^k(r)}{dr}\geq 0
\quad\Longleftrightarrow\quad
m'(r)\geq c^k(r)\cdot p'(r)$.
\end{corollary}

Here the term
\[
h^k(p(r),m(r))\cdot p'(r)
\]
is the first-order change in the cost of the current capitalist consumption bundle caused
by the change in prices as \(r\) rises. Therefore the condition
\[
m'(r)\geq h^k(p(r),m(r))\cdot p'(r)
\]
says that capitalist welfare is non-decreasing if and only if the change in capitalist
income is at least large enough to compensate capitalists for the price change evaluated
at their current consumption bundle.

In the numerical example capitalist preferences are Cobb--Douglas:
\[
U^k(c)=\prod_{i=1}^n c_i^{\beta_i},
\qquad
\beta_i>0,
\qquad
\sum_{i=1}^n\beta_i=1.
\]
Then the indirect utility function is
\[
V^k(p,m)
=
m\prod_{i=1}^n
\left(\frac{\beta_i}{p_i}\right)^{\beta_i}.
\]
Thus, along the equilibrium path,
\[
U^k(r)
=
m(r)
\prod_{i=1}^n
\left(\frac{\beta_i}{p_i(r)}\right)^{\beta_i}.
\]
Capitalist welfare is locally weakly increasing at \(r\) if and only if
\[
\frac{m'(r)}{m(r)}
\geq
\sum_{i=1}^n \beta_i \frac{p_i'(r)}{p_i(r)}.
\]
Using the equilibrium expression for \(m'(r)\), this is equivalent to
\[
\frac{
e'(r)-H_r(r,m(r))
}{
H_m(r,m(r))m(r)
}
\geq
\sum_{i=1}^n \beta_i \frac{p_i'(r)}{p_i(r)}.
\]

\subsection{Fixed consumption proportions and monotonic capitalist outcomes}\label{subsec:fixed}

The non-monotonicity results above depend on changes in the composition of consumption bundles.
It is therefore useful to isolate the benchmark case in which consumption proportions are
fixed. The following theorem shows that if workers and capitalists consume along a fixed ray, then total profits and capitalist welfare are
unambiguously increasing in the profit rate.

Let \(b^\ell\in\mathbb{R}^n_+\setminus\{0\}\) denote the fixed workers' consumption
basket, or consumption direction. Since workers' income is normalized to one and workers
exhaust their budget, workers' consumption is
\[
c^\ell(r)=\frac{b^\ell}{p(r)b^\ell}.
\]
Thus workers always consume goods in the same physical proportions.

Similarly, let \(b^k\in\mathbb{R}^n_+\setminus\{0\}\) denote the fixed capitalist
consumption direction. Capitalist consumption is assumed to be
\[
c^k(r)=\lambda(r)b^k
\]
for some scalar \(\lambda(r)>0\). This assumption can be interpreted either as fixed
capitalist consumption proportions or as the Marshallian demand generated by preferences
whose optimal bundle is always proportional to \(b^k\).

\begin{theorem}[Fixed proportions imply monotonic profits and capitalist welfare]
\label{thm:fixed_baskets_monotonicity}
\[
\frac{d}{dr}e(r)>0,
\]
\[
\frac{d}{dr}\pi(r)>0,
\]
and
\[
\frac{d}{dr}U^k(c^k(r))>0.
\]
Hence, under fixed workers' and capitalists' consumption proportions, the exploitation
rate, total profits, and capitalist welfare are all strictly increasing in the profit rate.
\end{theorem}

Why fixed workers' proportions alone are not enough?
The theorem requires fixed capitalist proportions as well as fixed workers' proportions.
Fixed workers' proportions alone imply
\[
e'(r)>0,
\]
because workers' labor-value consumption
\[
vc^\ell(r)=\frac{vb^\ell}{p(r)b^\ell}
\]
falls as \(p(r)b^\ell\) rises. However, this does not by itself guarantee that
\[
\pi(r)=p(r)c^k(r)
\]
is increasing.

The reason is that capitalist demand may change composition as \(r\) changes. In the
notation used above,
\[
H(r,m)=vh^k(p(r),m).
\]
Even when \(e'(r)>0\), equilibrium expenditure satisfies
\[
H(r,m(r))=e(r),
\]
so
\[
m'(r)=\frac{e'(r)-H_r(r,m(r))}{H_m(r,m(r))}.
\]
Thus, if capitalist substitution makes \(H_r(r,m(r))\) sufficiently positive, then
\(m'(r)\) may be negative even though \(e'(r)>0\).

Equivalently, fixed workers' proportions eliminate the possibility that workers' changing
composition makes the residual labor value fall. But they do not eliminate the possibility
that capitalists switch toward goods with a higher labor value per unit of expenditure.
Such a capitalist substitution effect can still make total profits decrease unless the
capitalist consumption direction is also fixed, or unless one imposes the weaker condition
\[
H_r(r,m(r))\leq 0.
\]


\section{Conclusion and implications for distributive conflict}\label{conclusion}

This paper has examined how consumption preferences can be incorporated into a Sraffian model of simple reproduction without abandoning the classical determination of prices and distribution. Production prices and the wage--profit configuration are determined independently of individual utility maximization. Once distribution is fixed, however, workers and capitalists choose consumption bundles subject to their class incomes and the corresponding prices of production. This allows us to define \textit{distributive Engel curves}, which describe how class-specific consumption bundles vary with the profit rate.

The main result is that the relation between the profit rate and capitalist outcomes need not be monotonic. In the fixed bundles or one-consumption-good case, the usual distributive intuition is confirmed: an increase in the profit rate reduces workers' consumption and increases capitalists' consumption. With more goods and varying consumption bundles, however, the outcome is more complex. A change in \(r\) affects consumption through two channels. It changes class incomes, but it also changes relative prices. For capitalists, these two effects may work in opposite directions. A higher profit rate may raise the capitalist share of the surplus while simultaneously increasing the relative price of the goods that capitalists prefer. In that case, capitalist real consumption and welfare may fall even as the profit rate rises.

The numerical example illustrates this possibility. Over a range of profit rates, workers' substitution toward goods with higher labor values reduces the residual labor value available for capitalist consumption and therefore reduces surplus value rate. As a consequence, total profits, measured as capitalist expenditure \(p(r)c^k(r)\), and capitalist welfare \(U^k(r)\) can both decline. The profit rate, total profits, and capitalist welfare are distinct objects. A higher profit rate does not necessarily imply higher total profits, and higher total profits do not necessarily imply higher welfare once relative-price effects are taken into account.

This result qualifies the standard interpretation of distributive conflict. The conflict between workers and capitalists remains central: in the model analyzed here, workers' welfare declines as the profit rate rises. Yet the analysis also reveals an additional contradiction within the capitalist class itself. Capitalists may collectively pursue a higher profit rate, but this pursuit can undermine their own consumption possibilities when the associated price changes are sufficiently unfavorable. In this sense, capitalist ``success'' in the distributive struggle may become self-defeating.

The analysis also has implications for the interpretation of the maximum profit rate. In the standard Sraffian framework, the maximum profit rate \(R\) is a technical magnitude determined by the input--output matrix. The present paper suggests that this technical bound need not coincide with the relevant bound from the standpoint of capitalist welfare. If capitalist welfare reaches a local maximum at some interior value \(r^*<R\), then further increases in the profit rate remain technically feasible but become welfare-reducing for capitalists themselves. Thus, a distinction should be drawn between the technically feasible maximum profit rate and the welfare-relevant range of profit rates.

More generally, the paper shows that introducing preferences into a classical framework does not require replacing production and distribution with subjective utility as the foundation of value and prices. Preferences enter at a different analytical level. They shape the composition of class consumption and the welfare consequences of distribution, while prices of production and the functional distribution of income remain classically determined. This separation makes it possible to study consumption behavior without collapsing the Sraffian framework into a marginalist one.

Several extensions are natural. Future work could consider non-homothetic preferences, alternative demand systems, joint production, fixed capital, or economies with government expenditure and foreign trade. Dynamic versions of the model could also study whether adjustments in consumption, prices, and distribution reinforce or dampen the non-monotonic effects identified here. Finally, the framework could be used to analyze how institutional constraints, social norms, or historically determined wage baskets affect the distributive Engel curves and the evolution of class welfare.

\bibliographystyle{econometrica}
\bibliography{utilidad}

\section*{Appendix: Mathematical proofs}

\section*{Proof of Theorem \ref{thm:necsuf_profit}}
\begin{proof}
Equilibrium requires
\[
H(r,m(r))=e(r).
\]
Differentiating both sides with respect to \(r\) gives
\[
H_r(r,m(r))+H_m(r,m(r))m'(r)=e'(r).
\]
Since \(H_m(r,m(r))\neq 0\), we can solve for \(m'(r)\):
\[
m'(r)
=
\frac{e'(r)-H_r(r,m(r))}{H_m(r,m(r))}.
\]

Because capitalist preferences are locally nonsatiated, capitalists exhaust
their budget:
\[
p(r)c^k(r)
=
p(r)h^k(p(r),m(r))
=
m(r).
\]
Therefore
\[
\frac{d}{dr}\bigl[p(r)c^k(r)\bigr]
=
m'(r).
\]
Hence
\[
\frac{d}{dr}\bigl[p(r)c^k(r)\bigr]\geq 0
\quad\Longleftrightarrow\quad
m'(r)\geq 0
\quad\Longleftrightarrow\quad
\frac{e'(r)-H_r(r,m(r))}{H_m(r,m(r))}\geq 0.
\]

If, in addition,
\[
H_m(r,m(r))>0,
\]
then the last inequality is equivalent to
\[
e'(r)-H_r(r,m(r))\geq 0,
\]
or
\[
e'(r)\geq H_r(r,m(r)).
\]
This proves the result.
\end{proof}

\section*{Proof of Theorem \ref{thm:necsuf_welfare}}
\begin{proof}
Equilibrium requires
\[
H(r,m(r))=e(r).
\]
Differentiating both sides with respect to \(r\) gives
\[
H_r(r,m(r))+H_m(r,m(r))m'(r)=e'(r).
\]
Since \(H_m(r,m(r))\neq 0\), we obtain
\[
m'(r)
=
\frac{e'(r)-H_r(r,m(r))}{H_m(r,m(r))}.
\]

Capitalist welfare is
\[
U^k(r)=W(r,m(r)).
\]
Differentiating gives
\[
\frac{dU^k(r)}{dr}
=
W_r(r,m(r))+W_m(r,m(r))m'(r).
\]
Substituting the expression for \(m'(r)\), we get
\[
\frac{dU^k(r)}{dr}
=
W_r(r,m(r))
+
W_m(r,m(r))
\frac{e'(r)-H_r(r,m(r))}{H_m(r,m(r))}.
\]
Therefore
\[
\frac{dU^k(r)}{dr}\geq 0
\]
if and only if
\[
W_r(r,m(r))
+
W_m(r,m(r))
\frac{e'(r)-H_r(r,m(r))}{H_m(r,m(r))}
\geq 0.
\]
This proves the first equivalence.

Multiplying numerator and denominator gives the equivalent expression
\[
\frac{
W_r(r,m(r))H_m(r,m(r))
+
W_m(r,m(r))\left[e'(r)-H_r(r,m(r))\right]
}{
H_m(r,m(r))
}
\geq 0.
\]
If \(H_m(r,m(r))>0\), this is equivalent to
\[
W_r(r,m(r))H_m(r,m(r))
+
W_m(r,m(r))\left[e'(r)-H_r(r,m(r))\right]
\geq 0.
\]
\end{proof}

\section*{Proof of Corollary \ref{cor:necsuf_roy}}
\begin{proof}
Since
\[
U^k(r)=V^k(p(r),m(r)),
\]
we have
\[
\frac{dU^k(r)}{dr}
=
\nabla_p V^k(p(r),m(r))\cdot p'(r)
+
V_m^k(p(r),m(r))m'(r).
\]
By Roy's identity,
\[
\nabla_p V^k(p(r),m(r))
=
- V_m^k(p(r),m(r)) h^k(p(r),m(r)).
\]
Therefore
\[
\frac{dU^k(r)}{dr}
=
V_m^k(p(r),m(r))
\left[
m'(r)-h^k(p(r),m(r))\cdot p'(r)
\right].
\]
Since \(V_m^k(p(r),m(r))>0\),
\[
\frac{dU^k(r)}{dr}\geq 0
\quad\Longleftrightarrow\quad
m'(r)\geq h^k(p(r),m(r))\cdot p'(r).
\]
Using
\[
m'(r)=\frac{e'(r)-H_r(r,m(r))}{H_m(r,m(r))},
\]
we obtain
\[
\frac{dU^k(r)}{dr}\geq 0
\quad\Longleftrightarrow\quad
\frac{e'(r)-H_r(r,m(r))}{H_m(r,m(r))}
\geq
h^k(p(r),m(r))\cdot p'(r).
\]
\end{proof}

\section*{Proof of Theorem \ref{thm:fixed_baskets_monotonicity}}
\begin{proof}
Since
\[
p(r)=\ell\left[I-(1+r)A\right]^{-1},
\]
differentiating gives
\[
p'(r)
=
\ell\left[I-(1+r)A\right]^{-1}
A
\left[I-(1+r)A\right]^{-1}.
\]
Because \(A\geq 0\), \(A\) is irreducible, \(\ell\gg 0\), and
\[
\left[I-(1+r)A\right]^{-1}\geq 0
\]
for \(r\in(0,R)\), we have
\[
p'(r)\gg 0.
\]
Therefore, for every nonzero \(b\in\mathbb{R}^n_+\),
\[
\frac{d}{dr}\bigl[p(r)b\bigr]=p'(r)b>0.
\]

Workers consume
\[
c^\ell(r)=\frac{b^\ell}{p(r)b^\ell}.
\]
Hence the labor value of workers' consumption is
\[
vc^\ell(r)
=
\frac{vb^\ell}{p(r)b^\ell}.
\]
Since \(vb^\ell>0\) is constant and \(p(r)b^\ell\) is strictly increasing in \(r\), it
follows that
\[
\frac{d}{dr}vc^\ell(r)<0.
\]
Therefore
\[
e(r)=1-vc^\ell(r)
\]
is strictly increasing:
\[
e'(r)>0.
\]

By the labor-value feasibility condition,
\[
v(c^\ell(r)+c^k(r))=1.
\]
Thus
\[
vc^k(r)=1-vc^\ell(r)=e(r).
\]
Since \(c^k(r)=\lambda(r)b^k\), we have
\[
\lambda(r)vb^k=e(r),
\]
and therefore
\[
\lambda(r)=\frac{e(r)}{vb^k}.
\]
Because \(vb^k>0\) and \(e'(r)>0\), it follows that
\[
\lambda'(r)>0.
\]
Thus capitalist physical consumption expands along the fixed ray \(b^k\).

Total profits are
\[
\pi(r)=p(r)c^k(r).
\]
Using \(c^k(r)=\lambda(r)b^k\), we obtain
\[
\pi(r)=\lambda(r)p(r)b^k.
\]
Differentiating,
\[
\pi'(r)
=
\lambda'(r)p(r)b^k+\lambda(r)p'(r)b^k.
\]
Now \(\lambda'(r)>0\), \(p(r)b^k>0\), \(\lambda(r)>0\), and \(p'(r)b^k>0\). Hence
\[
\pi'(r)>0.
\]
Therefore total profits are strictly increasing in \(r\).

Finally, capitalist welfare is
\[
U^k(c^k(r))=U^k(\lambda(r)b^k).
\]
Since \(\lambda'(r)>0\) and \(U^k\) is strictly increasing along the ray \(b^k\), it follows
that
\[
\frac{d}{dr}U^k(c^k(r))>0.
\]
This proves the theorem.
\end{proof}
\end{document}